%% file: main.tex
\documentclass[10pt,journal,compsoc]{IEEEtran}

\usepackage{booktabs,multirow}
\newcommand\Tstrut{\rule{0pt}{2.6ex}}    %
\newcommand\Bstrut{\rule[-0.9ex]{0pt}{0pt}} %

\usepackage{amsmath,amsfonts}
\usepackage{graphicx}
\usepackage{xcolor}

\usepackage{tikz}
\usetikzlibrary{shapes,arrows}

\usepackage{upquote}
\usepackage{fancyvrb}
\RecustomVerbatimEnvironment{Verbatim}{Verbatim}{fontsize=\small, frame=single, numbers=left, numbersep=2pt, xleftmargin=0.07in, commandchars=\\\{\}, tabsize=2}

\ifCLASSOPTIONcompsoc
  \usepackage[nocompress]{cite}
\else
  \usepackage{cite}
\fi

\ifCLASSOPTIONcompsoc
 \usepackage[caption=false,font=footnotesize,labelfont=sf,textfont=sf]{subfig}
\else
 \usepackage[caption=false,font=footnotesize]{subfig}
\fi
\usepackage{xurl}

\begin{document}

\title{ConfEx: A Framework for Automating Text-based Software Configuration Analysis in the Cloud}

\author{
       Ozan Tuncer,
       Anthony Byrne,
       Nilton Bila,
       Sastry Duri,
       Canturk Isci,
       and Ayse K. Coskun%
\IEEEcompsocitemizethanks{
\IEEEcompsocthanksitem
O. Tuncer, A. Byrne, and A. K. Coskun are with the Department of Electrical and Computer Engineering, Boston University, Boston, MA, 02215. \protect\\
E-mail: \{otuncer,abyrne19,acoskun\}@bu.edu
\IEEEcompsocthanksitem 
N. Bila, S. Duri and C. Isci are with IBM Research, T. J. Watson Research Center, Yorktown Heights, NY, 10598.  \protect\\
Email: \{nilton,sastry,canturk\}@us.ibm.com 
}%
}

\IEEEtitleabstractindextext{%
\begin{abstract} %

Modern cloud services have complex architectures, often comprising many software components, and depend on hundreds of configurations parameters to function correctly, securely, and with high performance.
Due to the prevalence of open-source software, developers can easily deploy services using third-party software without mastering the configurations of that software.
As a result, configuration errors (i.e., misconfigurations) are among the leading causes of service disruptions and outages.
While existing cloud automation tools ease the process of service deployment and management, support for detecting misconfigurations in the cloud has not been addressed thoroughly, likely due to the lack of frameworks suitable for consistent parsing of unstandardized configuration files.

This paper introduces \textit{ConfEx}, a framework that enables \textit{discovery and extraction} of text-based software configurations in the cloud.
\textit{ConfEx} uses a novel vocabulary-based technique to identify configuration files in cloud system instances with unlabeled content.
To extract the information in these files, \textit{ConfEx} leverages existing configuration parsers and post-processes the extracted data for analysis.
We show that \textit{ConfEx} achieves over 99\% precision and 100\% recall in identifying configuration files on 7805 popular Docker Hub images.
Using two applied examples, we demonstrate that \textit{ConfEx} also enables detecting misconfigurations in the cloud via existing tools that are designed for configurations represented as key-value pairs, revealing 184 errors in public Docker Hub images.

\end{abstract}

}

\maketitle

\input{intro}

\input{background}
\input{framework}

\input{evaluation}

\input{case_studies}

\input{related}

\input{conclusion}
\ifCLASSOPTIONcaptionsoff
  \newpage
\fi

\bibliographystyle{IEEEtran}
\bibliography{IEEEabrv,main}

\vspace{-0.3in}
\begin{IEEEbiographynophoto}{Ozan Tuncer}
is a Ph.D. candidate at the Department of Electrical and Computer Engineering of Boston University. He received his M.S. degree in Computer Engineering from Boston University, and his B.S. degree in Electrical and Electronics Engineering from the Middle East Technical University, Turkey. His research interests include data analytics for cloud system management, data center power and thermal management, and workload management
for high performance computing.
\end{IEEEbiographynophoto}

\vspace{-0.3in}
\begin{IEEEbiographynophoto}{Anthony Byrne}
is a Ph.D. student at the Department of Electrical and Computer Engineering of Boston University. He received his B.S. degree in Computer Engineering from Boston University. His current research interests include operating systems, embedded systems, and applications of machine learning in cloud systems. 
\end{IEEEbiographynophoto}

\vspace{-0.3in}
\begin{IEEEbiographynophoto}{Nilton Bila}
is a systems researcher at IBM. He builds large scale cloud platforms and  addresses problems in systems configurations, cloud security, software dependability, and virtualization. He received his Ph.D. and M.Sc. degrees in Computer Science from the University of Toronto.
\end{IEEEbiographynophoto}

\vspace{-0.3in}
\begin{IEEEbiographynophoto}{Sastry Duri}
is a senior software engineer at the IBM T. J. Watson Research Center in
Yorktown Heights, New York. He earned a Ph.D. degree in computer science from the University of Illinois at Chicago. His professional interests include distributed computing systems, cloud monitoring and analytics, mobile commerce applications, and RFID based supply chains. In his spare time, Sastry coaches teams for FIRST Robotics competitions. In the past, he represented IBM in the industry standard group EPCglobal ALE Working Group, a subsidiary of the Uniform Code Council (UCC), and in OpenLS workgroup.
\end{IEEEbiographynophoto}

\vspace{-0.3in}
\begin{IEEEbiographynophoto}{Canturk Isci}
is a principal researcher and master inventor in IBM T.J. Watson Research Center. He currently works on cloud monitoring, operational and security analytics. Prior to IBM Research, Canturk was a Senior Member of Technical Staff at VMware, where he worked on distributed resource and power management. Canturk has a B.S. in Electrical Engineering from Bilkent University, an M.Sc. with Distinction in VLSI System Design from University of Westminster, and a Ph.D. in Computer Engineering from Princeton University.
\end{IEEEbiographynophoto}

\vspace{-0.3in}
\begin{IEEEbiographynophoto}{Ayse K. Coskun}
is an associate professor at the Department of Electrical and
Computer Engineering, Boston University (BU). She was with Sun Microsystems (now Oracle), San Diego, prior to her current position at BU. Her research interests include energy-efficient computing, architectures built with emerging technologies, embedded systems, and large-scale systems analytics and management. She received the M.S. and Ph.D. degrees in Computer Science and Engineering from the University of California, San Diego. She currently serves as an associate editor of IEEE Transactions on CAD.
\end{IEEEbiographynophoto}

\vfill

\end{document}

%% file: intro.tex
\IEEEraisesectionheading{\section{Introduction}\label{sec:introduction}}

\IEEEPARstart{C}{loud} applications are designed in a highly configurable way to ensure high levels of reusability and portability.
To function correctly, securely, and with high performance, these applications often depend on precise tuning of hundreds of configuration parameters~\cite{Xu:FSE15:tooManyKnobs}.
The number of parameters to tune can reach thousands in typical cloud services that consist of multi-tiered software stacks~\cite{Ramachandran:ICAC09:dependencies}.

While configurations are traditionally validated by applications during startup, recent work has shown that, across various software applications in today's cloud, 14-93\% of configuration parameters do not have any special code for checking their correctness~\cite{Xu:OSDI16:earlyDetection}.
Moreover, the affordability offered by the cloud and the prevalence of open-source software have enabled new levels of agility, where small teams of developers can deliver new cloud services and functionality in short periods of time.
This newfound agility has led to a trend where service developers and operators leverage third-party software and public cloud images without necessarily having the expertise needed to precisely configure all components of their service. This, combined with the often-immature documentation that accompanies newly-introduced software, makes human error the leading cause of configuration-related failures~\cite{Xu:2015:SAT}. In a similar vein, configuration errors have taken their place among the leading causes of cloud software failures~\cite{oppenheimer:internet03:fail,Rabkin:software13:hadoop,Yin:SOSP11:ESC:empirical}, and have been reported as causes of service disruptions at Microsoft Azure~\cite{outage_Azure}, Amazon EC2~\cite{outage_Amazon}, and Google~\cite{outage_Google}.

Existing failure avoidance and mitigation mechanisms in the cloud (e.g., redundancy or recovery) are insufficient to handle configuration errors, as configurations tend to affect the entire cloud service rather than a single component such as a VM or a process~\cite{oppenheimer:internet03:fail,maurer:acm15:fail}.
Widely-used cloud deployment tools such as Chef~\cite{chef} and Ansible~\cite{Ansible} provide centralized management support for configurations.
These tools manage the configuration parameters that are related to deployment and scaling but  do not typically validate the parameters that determine functionality and performance.
Hence, there is a growing need for support to help analyze and validate software configurations in cloud platforms.

There are several challenges in applying automated configuration analysis in the cloud.
First, users often do not store their configurations in standard file system locations in cloud instances (i.e., images, VMs, and containers).
Especially in multi-tenant cloud platforms, where there is no platform-wide configuration management mechanism, one needs to \textit{discover} the locations of configurations in a cloud instance to perform analysis.
Second, configuration parameters of cloud software are often embedded in human-readable text files, where each software has its own custom file syntax.
For automated analysis, the information extracted from these files needs to be represented in a consistent format that allows validation and comparison of individual parameters.
Third, cloud instances typically contain multiple configuration files that are tuned for different use cases or software versions.
As some of these files are not actively used by the running applications, one needs to determine which configurations are \textit{active} in a cloud instance to avoid false positives while detecting configuration errors.
Finally, as many configuration parameters are related to the execution environment~\cite{Zhang:ASPLOS14:encore}, environmental information such as file access permissions should be collected from cloud instances.
To address the above challenges, one needs a framework to \textit{discover and extract} consistent configuration information from cloud instances with unlabeled content.

In this paper, we introduce \textit{ConfEx}, a novel software configuration analytics framework that enables robust analysis of text-based software configurations in the cloud.
\textit{Conf\-Ex} collects environmental information from the cloud instances, discovers configuration files of known applications in these systems, and parses the discovered files to produce consistent configuration data.
By itself, \textit{ConfEx} does not perform any validation on the configuration data it extracts. Instead, it enables the use of existing validation tools originally designed for key-value-based configurations (such as \textit{PeerPressure}~\cite{Wang:OSDI04:peerPressure} and \textit{Encore}~\cite{Zhang:ASPLOS14:encore}) on extracted configuration data. %
As we demonstrate in our evaluation, without \textit{ConfEx}, these tools have limited applicability in the cloud. 
Our specific contributions are as follows:
\begin{itemize}
\item We propose \textit{ConfEx}, a configuration analytics framework that enables the analysis of text-based software configurations in multi-tenant cloud platforms and image repositories. We provide two examples of \textit{ConfEx} being applied to existing configuration analysis tools to detect misconfigurations.%
\item As part of \textit{ConfEx}, we develop a method to discover the configuration files in cloud instances with unlabeled content. By identifying configuration keywords (such as parameter names) in an application-agnostic manner, our method achieves over 99\% precision and 100\% recall on identifying configuration files in 7805 Docker Hub images.
\item To enable focusing on configuration files that are actively used by running applications, we develop two methods: Our first method targets VMs and utilizes Linux file timestamps; our second method tracks system calls during application initialization and targets containers.
\item We demonstrate that the outputs of existing configuration parsers often lack the consistency and robustness for configuration analysis. To resolve this issue, we design a \textit{disambiguation} methodology, enabling comparison and analysis of configurations among thousands of cloud instances.
\end{itemize}

%% file: background.tex
\vspace{-0.15in}
\section{Background on Cloud Software Configurations}
\label{sec:background}

Most cloud applications and system services store their configurations in human-readable text files or in configuration stores such as \texttt{etcd} or Windows registry.
We focus on text file based configurations as this type of storage is prevalent for many of the building blocks of cloud applications (e.g., MySQL, Nginx, and Redis).
The remainder of this section explains cloud software configurations in detail and discusses how to use analytics to detect configuration errors.

\subsection{Text-based Configurations}
\label{sec:background:configs}

\begin{figure}[t]
\begin{Verbatim}[numberblanklines=false]

ServerRoot "/var/www"
Listen 80
<IfModule unixd_module>
    User daemon
    Group daemon
</IfModule>
\end{Verbatim}
\vspace{-1.12in}
\tikzset{>=latex}
\hspace{0.01in}
\begin{tikzpicture}[font=\sffamily\footnotesize]
  \node [gray, anchor=south] (parameter) at (1, 1.22) {parameter};
  \node [gray, anchor=south] (value) at (2.8, 1.23) {value};
  \node [gray, anchor=west, text width=3cm] (conditional) at (3.6,.9) {application-specific conditional statement};

  \node(first)[draw,ellipse,gray, minimum width=.8in, minimum height=.2in] at(-.2, 1) {};
  \node(second)[draw,ellipse,gray, minimum width=.8in, minimum height=.2in] at(1.9, 1) {};
  \node(third)[draw,ellipse,gray, minimum width=1.85in, minimum height=.2in] at(1.1, .3) {};

  \draw[gray, ->] (first.45) -- (parameter.west);
  \draw[gray, ->] (second.45) -- (value.west);
  \draw[gray, ->] (third.5) -- (conditional.west);
\end{tikzpicture}
\vspace{0.35in}
\caption{Httpd configuration file snippet. Configurations are stored in an XML-like format.}
\label{snippet:httpd}

\vspace{0.15in}
\begin{Verbatim}
proc   swap     swap   pri=42         0 0
tmpfs  /dev/shm tmpfs  mode=0777      0 0
devpts /dev/pts devpts defaults,gid=5 0 0
\end{Verbatim}
\vspace{-0.1in}
\caption{\texttt{/etc/fstab} snippet. Configurations are stored in a table format where certain table cells contain multiple configuration entries.}
\label{snippet:fstab}
\vspace{-0.1in}
\end{figure}

Figure~\ref{snippet:httpd} shows a snippet from an Apache HTTP server (httpd) configuration file.
Each of the first two lines contains a \textit{parameter} followed by a \textit{value}, separated by a space.
Lines 3-6 are in an application-specific format representing a conditional statement.
\texttt{User} and \texttt{Group} parameters are defined within the \texttt{IfModule unixd\_module} section, representing a configuration hierarchy.

In some configuration files, understanding the file schema %
requires domain knowledge. %
One such example is the Linux filesystem configuration file (\texttt{/etc/fstab}). %
As shown in Figure~\ref{snippet:fstab}, this file is structured in a table format where some columns may include parameter-value pairs such as \texttt{pri=42} (line 1) or multiple comma-separated entries such as \texttt{defaults,gid=5} (line 3).

Extracting configuration data from text-based files while retaining the relational information between different entries requires expertise on the specific file format.
Hence, to conduct corpus-based configuration analysis on a large number of applications, one should use a parsing tool that is continuously maintained by application domain experts.

\subsection{Configuration File Locations}
\label{sec:backgroud:location}

Software package managers such as \texttt{rpm} place configuration files to specific file system paths by default.
For example, the configurations of MySQL are installed by default to \texttt{/etc/my.cnf}, \texttt{/etc/mysql/my.cnf}, and \texttt{/etc/mysql/conf.d/}.

In cloud platforms, however, users often store their configurations in non-standard locations.
By examining popular Docker Hub images as an example, we have discovered that depending on the application, 20-81\% of configuration files are located in non-standard paths (see Sec.~\ref{sec:results:discovery} for details).
For example, we have identified MySQL configuration files located in \texttt{/app/my.cnf}, \texttt{/my-mini.cnf}, \texttt{/healthcheck.cnf}, and \texttt{/usr/cnf}, where file names are not necessarily indicative of MySQL.
Hence, one needs a systematic methodology to identify configuration files for comprehensive configuration analysis in the cloud.

\subsection{Active Configuration Files}
\label{sec:backgroud:active}

Cloud instances often contain multiple syntactically-valid configuration files for a given application.
These files include (1) configurations of modules and plug-ins, which enable new software functionalities, (2) configurations of specific application instances (e.g., Nginx can use a separate configuration file for each \textit{virtual host}), (3) template files, which set the default application behavior and can be directly used, and (4) multiple versions of configuration files that are used for testing and development purposes.

At the time of deployment, an application will use only a specific set of configuration files, which we refer to as \textit{active} configuration files.
These files are determined based on command line options or the main configuration file. The remaining \textit{passive} files typically do not affect the application behavior.
Detecting misconfigurations in these passive files would be of little use to cloud users and may be even seen as false positives.
Hence, it is necessary to identify which files are actively used in a running cloud instance %
for accurate error detection.

\subsection{Configuration Errors}
\label{sec:background:errors}

Table~\ref{tab:misconfig_types} summarizes common misconfiguration types we derived from related work (e.g.,~\cite{Ramachandran:ICAC09:dependencies,Yin:SOSP11:ESC:empirical,Zhang:ASPLOS14:encore,Xu:SOSP13:doNotBlame,Chen:COMPSAC16:correlation}) and online technical forums (e.g., \url{stackoverflow.com}).
\textit{Illegal entries} can be identified through syntactic validation. %
Detecting \textit{inconsistent entries} and \textit{invalid ordering} requires extracting dependency and correlation information among various parameters.
\textit{Environmental inconsistencies} occur when application configurations do not match the environmental parameters such as file permissions and IP addresses.
To find such inconsistencies, one needs to collect and analyze both application and environment configurations.
Detecting \textit{missing parameters} requires checking the existence of parameters rather than focusing on the values assigned to parameters.
\textit{Valid entries} that cause performance degradation or security vulnerabilities do not lead to crashes or error messages.

\begin{table}[t]
\centering
\setlength\tabcolsep{1.5pt}
\caption{Common configuration error types and example constraints that lead to errors upon violation.}
\label{tab:misconfig_types}
\vspace{-0.1in}
\begin{tabular}{cl}
\toprule
\begin{tabular}{c}Error type\end{tabular}                    & \multicolumn{1}{c}{Example configuration constraint}                                                                                                                                                                                                                 \\ \midrule
\multirow{2}{*}{Illegal entries}                                                   & 
\begin{tabular}{l} In PostgreSQL, parameter values that are not simple\\identifiers or numbers must be single-quoted.\end{tabular} \\ \cmidrule(l){2-2} 
& Variables must be in certain types (e.g., float). \\ \midrule
\multirow{2}{*}[-2em]{\begin{tabular}{c}Inconsistent\\entries\end{tabular}}
& \begin{tabular}{l} In PHP, \texttt{mysql.max\_persistent} must be no\\larger than the \texttt{max\_connections} in MySQL. \vspace{-0.05in}\end{tabular} \\ \cmidrule(l){2-2} 
& \begin{tabular}{l}In Cloudshare, service's \texttt{redis.host} entry\\(an IP address) must be a substring of Nginx's \\ \texttt{upstream.msg.server} entry (IP address:port).\end{tabular} \\ \midrule
\begin{tabular}{c} Invalid \\ ordering \end{tabular}
& \begin{tabular}{l} When using PHP in Apache, \texttt{recode.so} must\\be defined before \texttt{mysql.so}. \end{tabular} \\ \midrule
\multirow{2}{*}[-1em]{\begin{tabular}[c]{@{}c@{}}Environmental\\inconsistency\end{tabular}}
& \begin{tabular}{l} In MySQL, maximum allowed table size must be\\smaller than the memory available in the system \end{tabular} \\ \cmidrule(l){2-2} 
& \begin{tabular}{l} In httpd, Apache user permissions must be set\\correctly to enable file uploads for website visitors. \end{tabular} \\ \midrule
\begin{tabular}{c} Missing \\ parameter \end{tabular}
& \begin{tabular}{l} In OpenLDAP, a configuration entry must include\\\texttt{ppolicy.schema} to enable password policy. \end{tabular} \\ \midrule
\multirow{2}{*}[1em]{\begin{tabular}{c} Valid entries \Tstrut\\ that cause \\ performance\\or security\\issues \end{tabular}}
& \begin{tabular}{l} MySQL's \texttt{Autocommit} parameter must be set to\\ \texttt{False} to avoid poor performance under ``insert''\\intensive workloads. \end{tabular} \\ \cmidrule(l){2-2}
& \begin{tabular}{l} Debug-level logging must be disabled to avoid\\performance degradation.  \end{tabular} \\ \bottomrule
\end{tabular}
\vspace{-0.15in}
\end{table}

\subsection{Configuration Analysis}

Researchers have developed various tools to automatically check for errors in software configurations (e.g., \cite{Attariyan:OSDI12:xray,Behrang:ESEC15:usersBeware}).
Unlike the configurations found in cloud instances, these tools are mainly geared towards configurations that are represented as key-value pairs, where each key consistently corresponds to a specific configuration parameter.

Among automated configuration validation tools, statistical and learning-based techniques (e.g., \cite{Potharaju:VLDB15:confSeer,Wang:OSDI04:peerPressure,Santolucito:plang17:synthesizing}) have gained popularity as low overhead configuration checkers that can be applied in an application-agnostic manner.
These techniques use a corpus of configurations collected from working systems to infer configuration constraints or learn common patterns.
Then, configurations that violate the inferred constraints or deviate from the norm are identified as potential errors.
Such methods are appealing in practice because they do not require intrusive dynamic analysis or application instrumentation.

Cloud environments, where a large number of users deploy their customized applications, provide a unique opportunity for statistical and learning-based configuration analysis. However, training models for such techniques and using these models to validate configurations requires \textit{discovery} of configurations and \textit{extraction} of configuration parameters across large populations of installed applications.

\subsection{Potential Uses for a Cloud Configuration Analytics Framework}
As mentioned in the introduction, the cloud's meteoric rise has enabled new highs in developer productivity. Automation, in terms of both cloud operators automating their infrastructure and cloud users automating their application deployments, has been a key driver of this rise~\cite{beyer2016site, devops}. 
However, a consistent method for automating cloud configuration analysis and validation has not yet been realized, and unless the right application-specific configuration validation tool happens to exist, service reliability engineers are often forced to manually review every revision of each application configuration that they manage.

A cloud-focused configuration analysis framework could lay the groundwork for a solution to this problem.
With the permission of cloud users, cloud operators could collect user configurations in VMs and containers to create a large and highly-diverse configuration dataset and to train accurate configuration models. These models could then be used to identify problematic configurations before deployment, saving cloud users days of engineering time that would otherwise be spent on configuration debugging.

%% file: framework.tex
\section{Configuration Analytics with ConfEx}
\label{sec:framework}

Our goal is to systematically analyze text-based software configurations in image repositories and multi-tenant cloud platforms, where cloud instances are calibrated by different users and include unlabeled content.
To this end, we design a configuration analytics framework, \textit{ConfEx}.

\begin{figure}
\centering
\includegraphics[width=0.85\columnwidth]{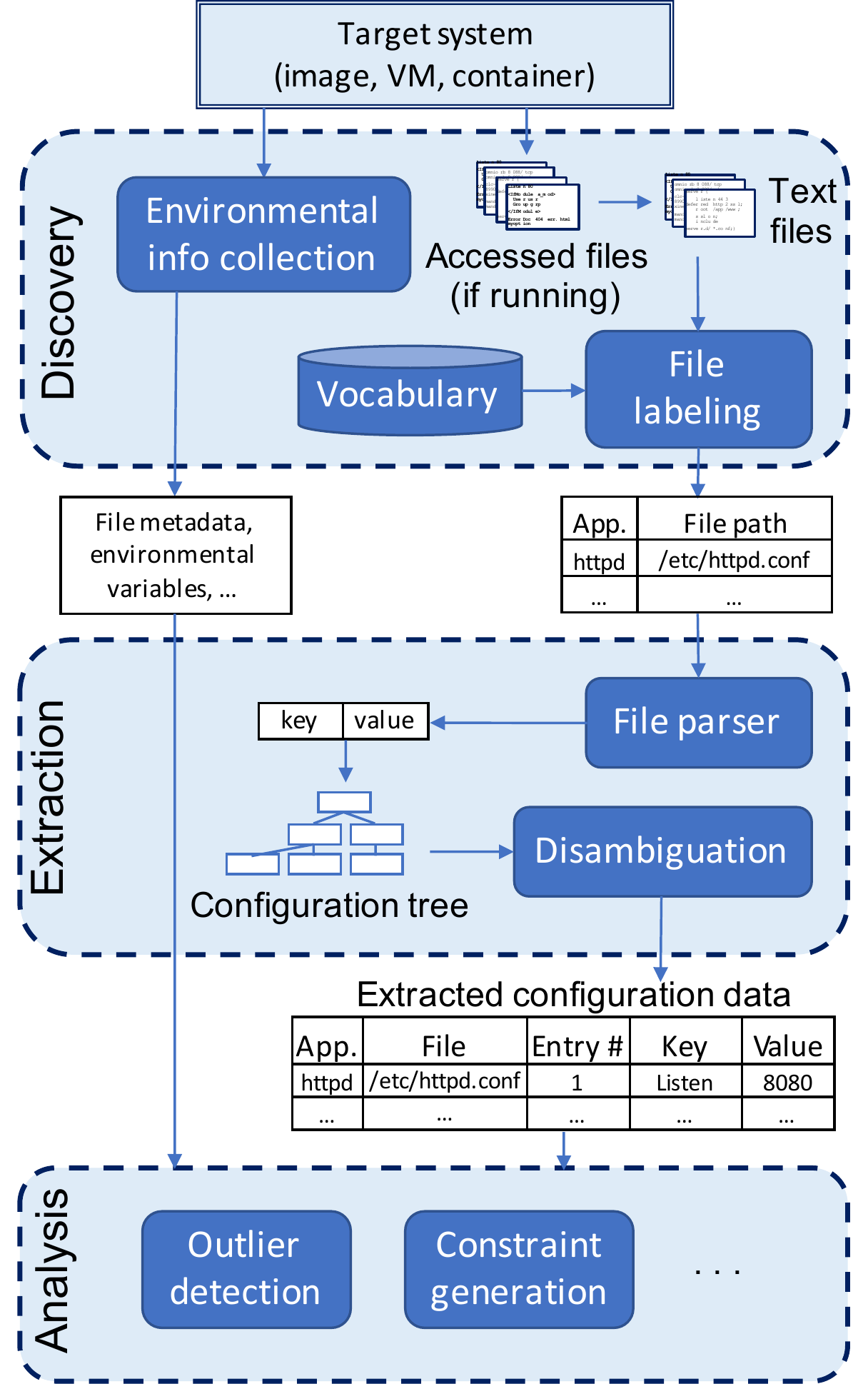}
\vspace{-0.05in}
\caption{\textit{ConfEx} overview.
The discovery phase collects environmental data and identifies actively used configuration files.
The extraction phase parses the information in the identified files and generates configuration data that is consistent across cloud instances.
The analysis phase applies existing rule-based, statistical, or learning-based tools to analyze and validate the configurations.}
\label{fig:framework}
\vspace{-0.1in}
\end{figure}

Figure~\ref{fig:framework} shows the overview of our \textit{ConfEx} framework.
\textit{ConfEx} has three phases: {\em discovery, extraction}, and {\em analysis}.
In the discovery phase, we identify the actively used configuration files and collect environmental data from target systems.
The extraction phase then parses the information in the configuration files, and transforms the collected data into key-value pairs where each key corresponds to a single configuration parameter.
Finally, in the analysis phase, we use existing tools based on outlier detection and rule-based validation to analyze the extracted configuration data.
The rest of this section explains these three phases in detail.

\subsection{Discovery}
Discovery focuses on (1) identifying configuration files that are actively used in the target system and (2) collecting environmental data (such as user privileges and file permissions) that can be used to detect misconfigurations.

\subsubsection{Active File Discovery}
As discussed in Sec.~\ref{sec:backgroud:active}, only a specific set of configuration files are used in a running cloud instance.
Including the remaining unused files in configuration analysis can lead to unreliable results, and misconfigurations detected in these unused files may be seen as false positives by cloud users.
Hence, \textit{ConfEx} focuses only on the files that are accessed by applications in VMs and containers.

We have developed two solutions to determine the files that are accessed: (1) checking file timestamps, and (2) tracking system calls during application initialization.
Our solutions, as described below, differ in their degree of applicability and intrusiveness.

\textbf{Checking file timestamps:}
One way to understand whether a file has been accessed is to check its Unix access timestamp, \texttt{atime}.
In most file systems, \texttt{atime} is updated by default when the file is read for the first time after being modified or if the existing \texttt{atime} is older than one day.
Hence, during discovery, we check \texttt{atime}'s and ignore the files that have not been accessed since a specific time point, which is the last system restart time by default and can be overriden by the user.
We use \texttt{atime} for active file discovery in VMs; however, this approach is not applicable if \texttt{atime} is disabled through file system mount options or in copy-on-write file systems such as shared Docker layers or \textit{btrfs}~\cite{rodeh:tos13:btrfs}.

\textbf{Tracking system calls:}
In systems where the \texttt{atime}-based approach is inapplicable, we identify accessed files by tracking \texttt{open()} system calls that have the \texttt{read} flag using the \texttt{auditd} tool on the host machine. 
Note that applications typically read their configurations during initialization; hence, we need to track system calls only during application initialization.
This approach is suitable for containers, which typically start with an entry script that runs the application inside the container. In situations where the application being analyzed is known to load configuration files long after its initialization (e.g., a modular web server), the user may configure ConfEx to continue monitoring syscalls for a longer duration.

Both active file discovery solutions are only applicable to cases where the target systems are actively running. %
In cases where the target system or image must be analyzed offline, active discovery can be skipped, and instead all available files can be passed to the identification stage.

\subsubsection{Configuration File Identification}
\label{sec:framework:file_identification}

The files identified by the active file discovery step include binaries and data files along with configuration files.
To identify the configuration files among the accessed files, we first discard non-text files. %
Second, to reduce computational overhead, we discard the files with extensions that are used for non-configuration files (such as \texttt{.h} or \texttt{.md5sums}) and the files that are larger than an empirically-determined size threshold (see Sec.~\ref{sec:evaluation} for details).
To identify configuration files among the remaining text files, \textit{ConfEx} examines the content of these files in the \textit{file labeling} step in Fig.~\ref{fig:framework}.

\begin{figure}
\centering
\includegraphics[width=0.92\columnwidth]{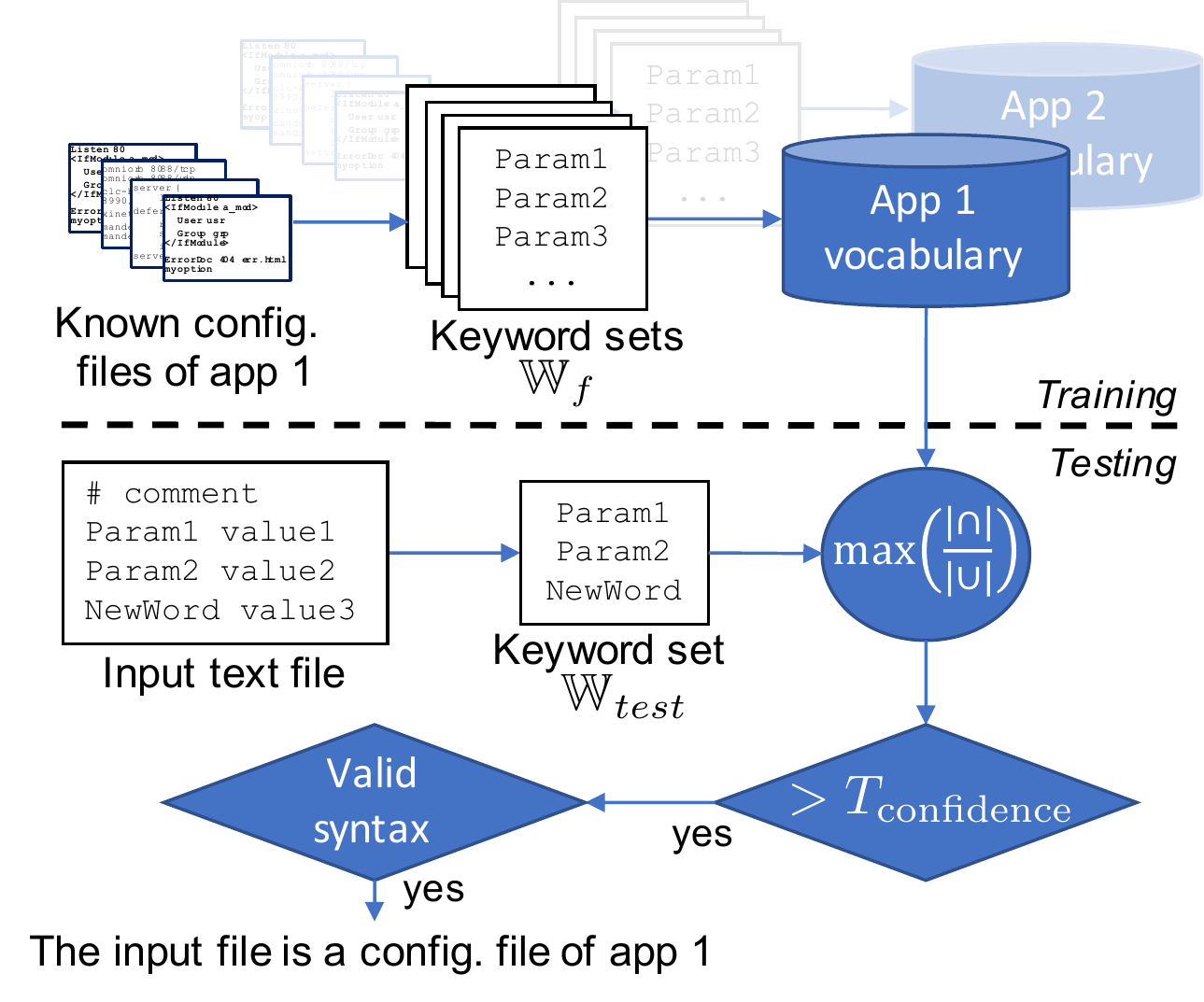}
\vspace{-0.1in}
\caption{File labeling step of the discovery phase. During offline training, a vocabulary is generated for each application using known configuration files. Input text files are compared with each application vocabulary. Upon a match that is larger than a confidence threshold, syntactically valid files are labeled as configuration files.}
\label{fig:framework:discovery}
\vspace{-0.15in}
\end{figure}

Figure~\ref{fig:framework:discovery} depicts \textit{ConfEx}'s file labeling step in detail.
During offline training, we use known configuration files that are labeled with application names.
We then identify the \textit{configuration keywords} in these files to generate application-specific vocabularies.
Configuration keywords include parameter names and configuration commands, and are usually specific to applications.
We extract these keywords in an application-agnostic way as follows:
We first discard commented-out lines, i.e., lines that begin with \texttt{//}, \texttt{\#}, or \texttt{\%}, excluding the preceding white-space characters.
The first words of non-comment lines in a configuration file typically correspond to parameter names or configuration commands, whereas the subsequent words are user-provided values such as integers and file paths.
Hence, we use the first word of the remaining lines as keywords.
While extracting keywords, we use the following characters as delimiters to account for the characters that are commonly used as part of a configuration file syntax: \texttt{\textbackslash t}, \texttt{=}, \texttt{ }, \texttt{:}, \texttt{<}, \texttt{>}, \texttt{[}, \texttt{]}, and \texttt{,}.
An application vocabulary consists of sets of unique keywords for each known configuration file.

During testing, we again extract the keyword set in the input text file using the same methodology.
We calculate the similarity of the input keyword set to each keyword set in the vocabulary of each application.
To calculate the similarity of a set pair, we use the Jaccard index~\cite{jaccard_index}, defined as $J=|\mathbb{W}_1 \cap \mathbb{W}_2| / |\mathbb{W}_1 \cup \mathbb{W}_2| $, where $\mathbb{W}_1$ and $\mathbb{W}_2$ are two sets.
If the maximum achieved similarity using the keyword sets in an application vocabulary is lower than an empirically-determined threshold, $T_{confidence}$, the file is discarded (see Sec.~\ref{sec:results:discovery} for details).
Finally, we check the syntax of the files with sufficient keyword set similarity, and label syntactically valid files as application configuration file.

Note that calculating the Jaccard index between the input keyword set, $\mathbb{W}_{test}$, and keyword sets for all known configuration files is computationally expensive.
Furthermore, most non-configuration files do not contain any application-specific keywords and need not to be compared with all keyword sets in a vocabulary.
Hence, we speed-up set comparison as follows:
Let a keyword set of a known configuration file $f$ be $\mathbb{W}_{f}$, and the union of all keywords in a vocabulary be $\mathbb{W}_{vocab}$.
Then, the Jaccard similarity ($J$) has the following upper bound:
\vspace{-0.1in}

\begin{equation}
J=\frac{\left| \mathbb{W}_{test} \cap \mathbb{W}_f \right|}{\left| \mathbb{W}_{test} \cup \mathbb{W}_f \right|} \leq J_{upper} = \frac{\left| \mathbb{W}_{test} \cap \mathbb{W}_{vocab} \right|}{\left| \mathbb{W}_{test} \right|}
\end{equation}

\noindent as %
$\mathbb{W}_f \subseteq \mathbb{W}_{vocab}$, and $\left| \mathbb{W}_{test} \cup \mathbb{W}_f \right| \geq \left| \mathbb{W}_{test} \right|$.
Checking $J_{upper}$ once per vocabulary eliminates the need to compare $\mathbb{W}_{test}$ with all $\mathbb{W}_f$'s in the vocabulary if $J_{upper} < T_{confidence}$.

To add a new application or extend an application's existing vocabulary, one can simply process new configuration files that are labeled with application names without the need of re-processing all known configuration files.

\subsubsection{Collecting Environmental Data}

As mentioned in Sec.~\ref{sec:background:errors}, misconfigurations can occur due to a mismatch between software configurations and environmental settings such as IP addresses and file permissions.
Table~\ref{tab:environmental_info} lists the environmental information \textit{ConfEx} collects.
All the information we collect can be used for analysis on VMs and containers; all except the network address and the active port information can be used for analysis on images.

\begin{table}[t]
\centering
\caption{Environmental information collected from cloud instances.}
\label{tab:environmental_info}
\vspace{-0.05in}
\begin{tabular}{cc}
\toprule
Description & Source\\
\hline
User information & \texttt{/etc/passwd} \Tstrut \\
Group information & \texttt{/etc/group} \\
File metadata & Crawling the file system \\
Environmental variables & \texttt{docker inspect} or \texttt{env} \\
Network addresses & \texttt{docker inspect} or \texttt{ifconfig} \\
Active ports & \texttt{docker inspect} or \texttt{netstat} \\
\bottomrule
\end{tabular}
\vspace{-0.15in}
\end{table}

\subsection{Extraction}

The extraction phase parses the configuration data located in text files identified by the discovery phase, and generates key-value pairs that represent configurations.
Such key-value pairs can be directly used by the existing configuration analysis tools such as Encore~\cite{Zhang:ASPLOS14:encore} and ConfigV~\cite{Santolucito:plang17:synthesizing}.

While existing studies on configuration analysis have mostly focused on configuration stores that do not require data extraction such as Windows Registry (e.g.,~\cite{Yuan:ATC11:contextBased}), or configurations with standard file formats such as \texttt{XML} or \texttt{JSON} (e.g.,~\cite{Behrang:ESEC15:usersBeware,Zhang:ISSTA15:diagnosticMessages}), most configuration files of cloud applications are kept in human-readable text files that do not use standard file formats.
As discused in Sec.~\ref{sec:background:configs}, these files require custom parsing rules based on domain knowledge.
However, the variety and rapid evolution of applications make it expensive and bug-prone to implement and maintain custom parsers for different applications for every configuration analysis tool.

\subsubsection{Augeas for Parsing Configuration Files}
\label{sec:framework:extraction:augeas}

To leverage the knowledge of domain experts on various applications and re-use an existing code-base that is continuously maintained, we build our extraction phase on top of Augeas~\cite{Augeas}, which is one of the most popular tools for automatized configuration parsing and editing.
Augeas has extensive application coverage with 182 \textit{lenses}, which are file parsing rules to generate key-value pairs for different applications including httpd, MySQL, Nginx, and PostgreSQL.
Augeas has been maintained for more than ten years, has interfaces in different programming languages including Python, Ruby, and Java, and used by other configuration management tools including Puppet~\cite{puppet} and bcfg2~\cite{bcfg2}.

As Augeas is primarily intended for managing configurations in systems with uniform and known configuration structure, its output is not ideal for key-value-based statistical analysis and learning in the cloud.
The issues with the Augeas parser output can be seen in the example in Fig.~\ref{fig:framework:extraction}.
Augeas produces artificial keys (e.g., \texttt{/directive[1]}) that do not correspond to parameters but represent the location and type of the configuration entries.
Hence, a specific Augeas key does not necessarily point to the same parameter across different files.
For example, in the httpd configuration file in Fig.~\ref{fig:framework:extraction}, if the first two lines were swapped, \texttt{/directive[1]} and \texttt{/directive[2]} keys would have referred to \texttt{Redirect} and \texttt{Listen}, respectively, unlike the Augeas output in Fig.~\ref{fig:framework:extraction}.
Because of this \textit{ambiguity} of Augeas key-value pairs, %
directly using Augeas is often ineffective for corpus-based configuration analysis.

\subsubsection{Disambiguation of the Augeas Output}
\label{sec:framework:disambiguation}

\begin{figure}[t]
\centering
\includegraphics[width=0.88\columnwidth]{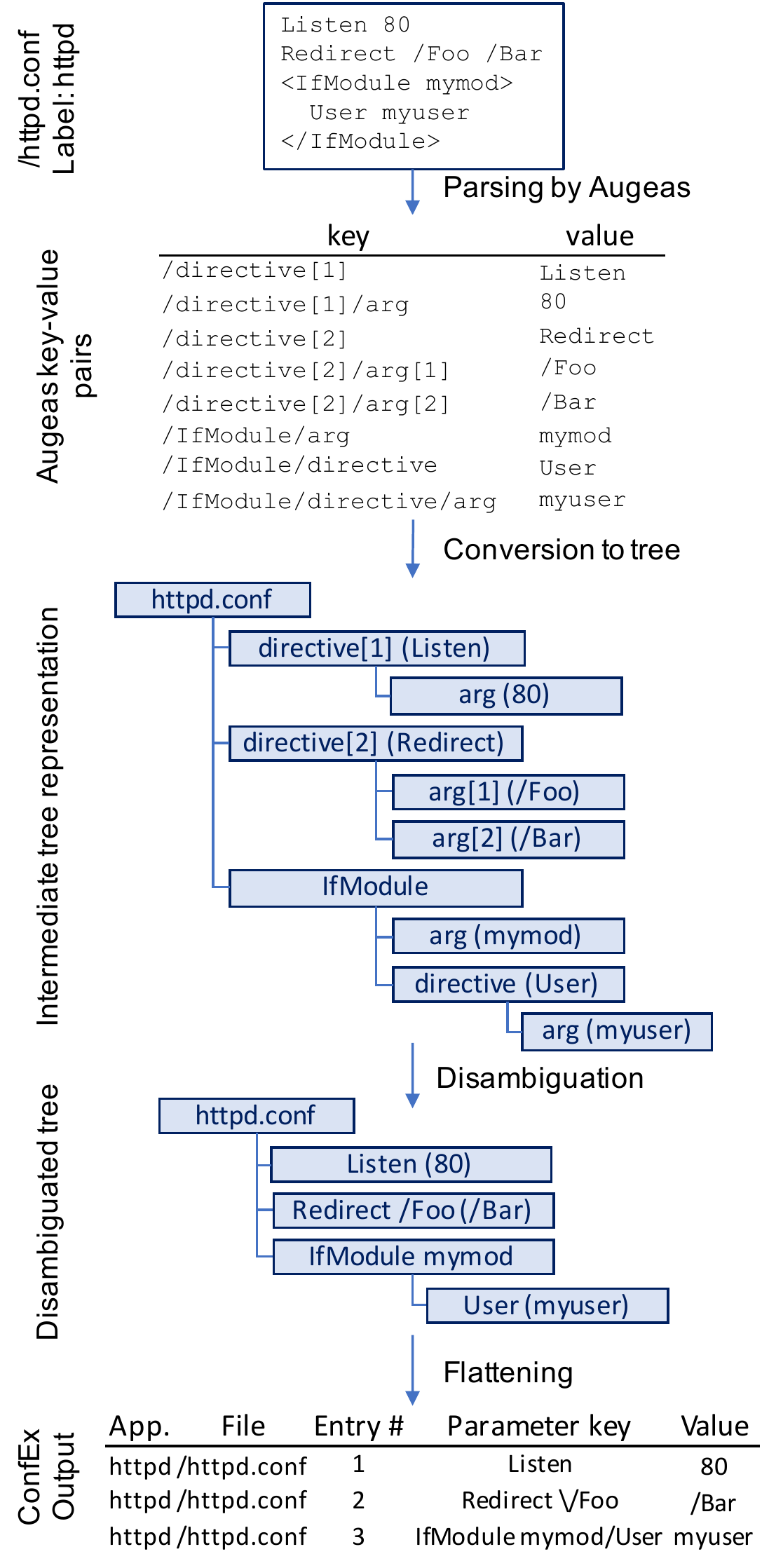}
\vspace{-0.15in}
\caption{Extraction phase. Augeas parses configuration files based on the labels given by the discovery phase. The key-value pairs generated by Augeas is converted into a tree that retains the configuration hierarchy, where texts in parenthesis represent the values of nodes. This tree is \textit{disambiguated} based on application-specific rules that are manually generated using minimal domain knowledge. The flattened form of the disambiguated tree contains key-value pairs where a key corresponds to a single parameter consistently across different files.}
\label{fig:framework:extraction}
\vspace{-0.15in}
\end{figure}

To prepare Augeas' output for corpus-based analysis, we use a \textit{disambiguation} step and transform Augeas' output into key-value pairs where a key consistently corresponds to the same single parameter across different files.
As depicted in Fig.~\ref{fig:framework:extraction}, %
we convert Augeas's output into an \textit{intermediate tree} that retains configuration hierarchy. %
We transform this tree using a list of application-specific rules such that the transformed tree faithfully represents all configuration parameters.
We manually implement these rules using minimal domain knowledge and only by examining the document structure, parameters found in the configuration files, and their corresponding output produced by Augeas.

\textbf{Example disambiguation rules:}
By examining httpd configuration files and the Augeas output, we observe that \texttt{directive} keys are redundant, and we extract the actual parameter names from the values of the \texttt{directive} keys.
The configuration options assigned to these parameters are extracted from the value of the child node named \texttt{arg}.
We also observe that specific entries such as \texttt{Redirect} %
represent configuration commands with multiple arguments.
From a configuration analysis perspective, we are interested in which arguments are being redirected (\texttt{/Foo} in Fig.~\ref{fig:framework:extraction}) and where they are directed to (\texttt{/Bar} in Fig.~\ref{fig:framework:extraction}).
In this case, we use \texttt{Redirect /Foo} as the key, indicating that \texttt{/Foo} is being redirected, and \texttt{/Bar} as the value assigned to this key.
We identify 15 such configuration commands in httpd documentations.
Our final observation is that nodes without values (such as \texttt{IfModule}) indicate configuration hierarchy.
These observations can be summarized in the following transformation rules for httpd:

\begin{itemize}
  \item \texttt{directive} nodes are replaced by the parameter names stored in the node's value. The value of the new node is the value of the child node named \texttt{arg}.
  \item For specific keys that represent configuration commands (such as \texttt{Redirect}), the new key is appended with the value of the child node named \texttt{arg[1]}. The value of the new node is the concatenation of the values of the remaining children whose name start with \texttt{arg}.
  \item Nodes without values (such as \texttt{IfModule}) are converted into an intermediate node where their key is appended with the value of the concatenation of the values of the children whose name start with \texttt{arg}.
\end{itemize}

After this rule-based transformation, the disambiguated tree is flattened and converted into a table as depicted in Fig.~\ref{fig:framework:extraction}.
In this table, the entry number represents the ordering of the values in the configuration file.
The application label and the file path are also appended to this table such that all configurations extracted from a cloud instance are represented in a single standardized format for analysis.

To extract reliable key-value pairs from the configuration files of a new application, one needs to implement tree transformation rules specific to the new application by examining the configuration file structure, configuration parameters, and the corresponding Augeas output using minimal domain knowledge as described above.
A new Augeas lens may be required if the Augeas library does not support the new application.

\vspace{-0.1in}
\subsection{Analysis}

The discovery and extraction phases of \textit{ConfEx} produce consistent key-value pairs that represent software configurations along with environmental information from the cloud instances, enabling the use of a rich variety of configuration analysis and validation techniques in multi-tenant cloud platforms and image repositories.
Analysis of software configurations can be used both to detect misconfigurations and to gain insight on user configuration practices.
Existing automated misconfiguration detection techniques that can be applied as part of \textit{ConfEx} include outlier value detection~\cite{Wang:OSDI04:peerPressure}, parameter type inference~\cite{Zhang:ASPLOS14:encore,Li:EASE17:confTest}, rule-based validation~\cite{Huang:EUROSYS15:confValley,Baset:Middleware17:declarative}, parameter correlation analysis~\cite{Chen:COMPSAC16:correlation}, matching configuration parameters with the parameters in the source code for source-based analysis~\cite{Zhou:EASE17:easier}. We provide examples of how two of these techniques could be used during this phase in Sec.~\ref{sec:case}. %

\vspace{-0.1in}
\subsection{Implementation}

We have implemented our \textit{ConfEx} framework using Python.
\textit{ConfEx} crawls cloud instances using IBM's public agentless system crawler\footnote{\url{https://github.com/cloudviz/agentless-system-crawler}} and uses Augeas 1.7 for file parsing.
We have implemented disambiguation rules for httpd, MySQL, and Nginx applications as well as \texttt{/etc/services}, \texttt{/etc/passwd}, and \texttt{/etc/group} system configurations.

%% file: evaluation.tex
\section{Evaluation}
\label{sec:evaluation}

We evaluate \textit{ConfEx} using the Docker Hub repository, which is one of the largest publicly available container image repositories with over 650,000 registered users~\cite{docker_users}.
We focus on the Docker Hub images that are either among the most downloaded 2000 images or contain one of the three following popular cloud applications: httpd, MySQL, and Nginx.
For each application, we use the images that are downloaded at least 50 times and contain the application name in their name or description.
We have manually labeled the configuration files in these images by examining file contents and paths of all text files that comply with the application configuration file syntax.
In addition to application configuration files, we use the following system configuration files in our evaluation: the network services file (\texttt{/etc/services}), the users file (\texttt{/etc/passwd}), and the groups file (\texttt{/etc/group}).

Table~\ref{tab:corpus} summarizes the number of images we use along with the number of identified configuration files in these images.
In total, we use 7805 images, where 1163 images contain configuration files of more than one target application and 254 images contain configuration file of all three applications.
The largest three configuration files have the sizes 140KB, 99KB, and 41KB.
Hence, we set the file discovery size threshold in the configuration file identification step (Sec.~\ref{sec:framework:file_identification}) to 200KB.
Our data set contains a total of over 22 million text files that are smaller than 200KB.

\begin{table}[t]
\centering
\caption{Statistics on the studied Docker Hub Images}
\label{tab:corpus}
\vspace{-0.1in}
\begin{tabular}{ccc}
\toprule
\begin{tabular}[c]{@{}c@{}} target\\application \end{tabular} & \begin{tabular}[c]{@{}c@{}} \# of images that\\contain the app. \end{tabular} & \begin{tabular}[c]{@{}c@{}} total \# of app.\\config. files \end{tabular} \Bstrut \\
\hline
httpd & 1601 & 53100 \Tstrut\\
MySQL & 2238 & 9481  \\
Nginx & 3714 & 32343 \\
Network services & 6106 & 6106 \\
Users \& Groups & 7805 & 7805 \\
\bottomrule
\end{tabular}
\vspace{-0.15in}
\end{table}

The remainder of this section first discusses our findings based on our study of the active files in Docker containers.
We then study the impact of $T_{confidence}$ on configuration file discovery, compare \textit{ConfEx}'s discovery phase with the baseline approaches, and discuss the overhead of \textit{ConfEx}.

\begin{figure*}
\subfloat[httpd\label{fig:discovery:threshold:httpd}]{
\includegraphics[width=0.34\textwidth]{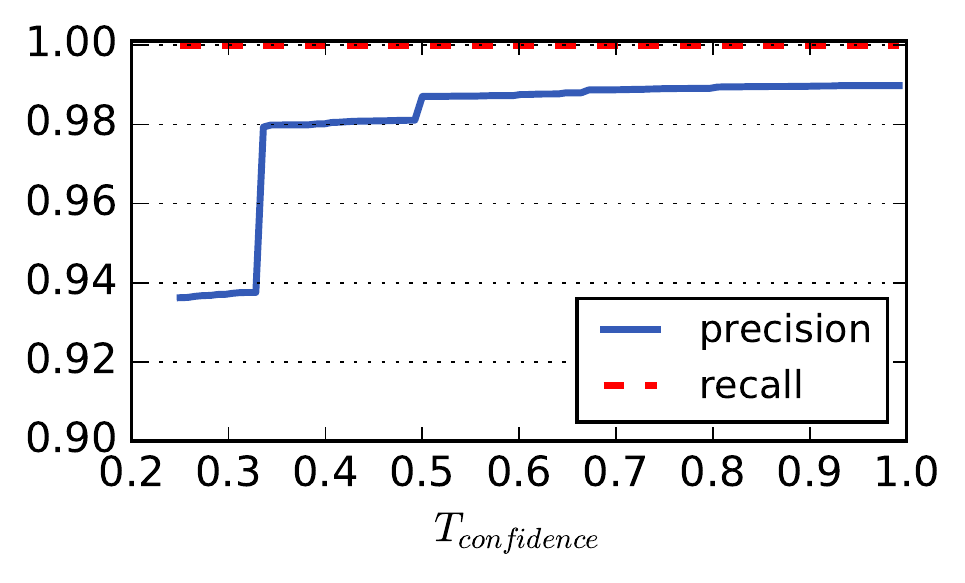}
}
\subfloat[MySQL\label{fig:discovery:threshold:mysql}]{
\includegraphics[width=0.33\textwidth]{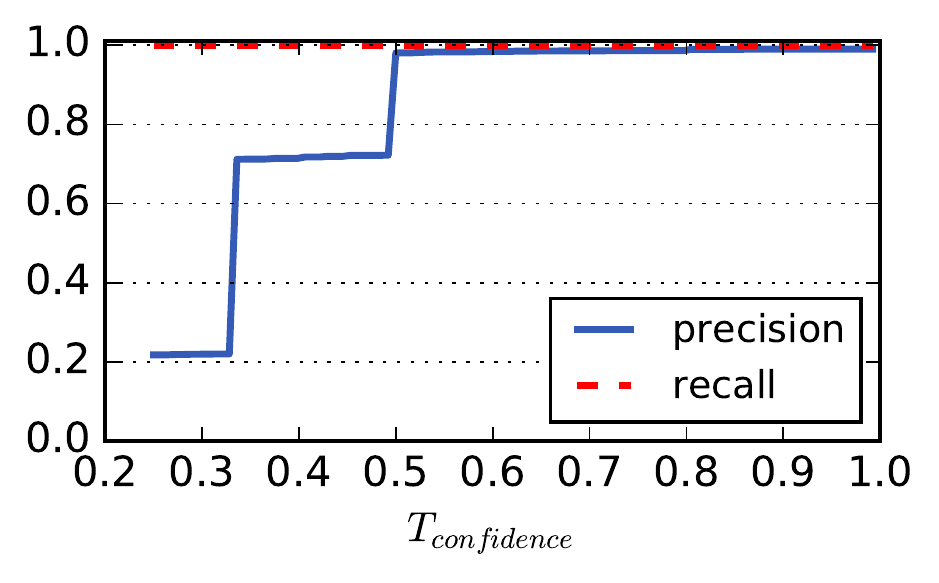}
}
\subfloat[Nginx\label{fig:discovery:threshold:nginx}]{
\includegraphics[width=0.33\textwidth]{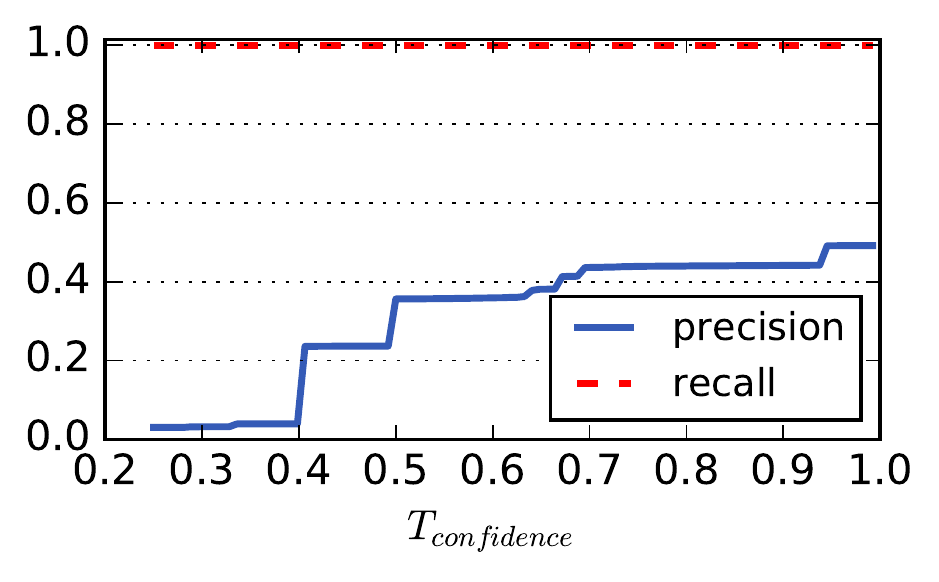}
}
\vspace{-0.05in}
\caption{Configuration file identification results w.r.t. confidence threshold using only the vocabulary based identification (without syntax check). Recall remains nearly ideal for all confidence thresholds as configuration files use the same set of commands and parameter names. With a confidence threshold above 0.5, \textit{ConfEx} achieves above 0.98 precision for httpd and MySQL even without syntax check.}
\label{fig:discovery:threshold}
\vspace{-0.1in}
\end{figure*}

\subsection{Active File Discovery}

\begin{table}
\centering
\caption{Statistics on the active configuration files in the selected Docker Hub images}
\label{tab:active}
\vspace{-0.1in}
\begin{tabular}{cccc}
\toprule
application & \begin{tabular}[c]{@{}c@{}} \# of\\images \end{tabular} &  \begin{tabular}[c]{@{}c@{}} \# of accessed\\ files \end{tabular} &  \begin{tabular}[c]{@{}c@{}} \# of accessed\\ config. files \end{tabular} \Bstrut \\
\hline
httpd & 50 & 2415  (out of 559K) & 496 (out of 1768) \Tstrut\\
MySQL & 50 & 10152 (out of 1.1M) & 143 (out of 290)  \\
Nginx & 50 & 5788  (out of 404K) & 166 (out of 650) \\
\bottomrule
\end{tabular}
\vspace{-0.15in}
\end{table}

To identify accessed (i.e., active) files in Docker containers, \textit{ConfEx} tracks \texttt{open()} system calls to the container file system during application initialization.
In our experiments, we find that tracking \texttt{open()} calls for ten seconds is sufficient to capture configuration file accesses of the applications that are already installed in the images.
After ten seconds, we observe either no more \texttt{open()} calls or periodic calls to specific non-configuration files.

The arguments used while deploying a Docker container may affect which configuration files are used by the running applications.
However, most images on Docker Hub do not contain instructions on their intended deployment procedure.
Hence, we study the \textit{active file discovery} phase of \textit{ConfEx} using a subset of our target images where we use the necessary commands to correctly initialize the applications.

Table~\ref{tab:active} shows statistics on the active files in the selected Docker Hub images.
On average, less than 1\% of the existing files are accessed during application initialization.
This indicates that most Docker images are loaded with extra files that are not needed by the applications.
We find that while using these publicly available images significantly reduces development time, these images often contain files and features that are not needed by the final service, filling the deployed container with unnecessary files.
For example, the \texttt{ubuntu:xenial} image, which a na\"ive developer may use as a base image, contains over 1000 user manual files, which are not used in automatically deployed containers, and over 500 files for the \texttt{perl} package, which may not be needed by the running applications.

Our results in Table~\ref{tab:active} also show that only 30\% of the syntactically valid configuration files are used by the running applications.
As discussed in Sec.~\ref{sec:background:configs}, the remaining configuration files consist of configurations of unused modules and plug-ins, template files, and configuration files that are used for testing and development purposes rather than for deployment.
Errors detected in these unused files can be perceived as false positives by cloud users.
\textit{ConfEx} prevents such false positives by focusing on active files in VMs and containers.
However, for static images, no file access information is available.
Thus, configuration analysis should be performed on all configuration files in images.

\subsection{Configuration File Identification}
\label{sec:results:discovery}

We measure the effectiveness of configuration file identification separately for each application and using five-fold cross validation.
That is, for each application, we randomly divide the images in our corpus into five equal-sized partitions.
We use the configuration files of the target application in four of these partitions to train our framework, and all the text files of the fifth partition as testing set, where we predict whether the input text files are configuration files of the target application.
We repeat this procedure five times, where each partition is used as a testing set once.
Furthermore, we repeat the five-fold cross validation five times with different randomly-selected partitions.

We use \textit{precision} and \textit{recall} as evaluation metrics. 
Precision is the fraction of true positives (i.e., correctly identified configuration files) to the total number of files predicted as configuration files, and recall is the fraction of true positives to the total number of configuration files in the testing set.

\subsubsection{Selecting the Confidence Threshold}

Figure~\ref{fig:discovery:threshold} shows the precision and recall \textit{ConfEx} achieves on identifying the configuration files with various $T_{confidence}$ levels only using the vocabulary-based identification (i.e., without syntax check).
Recall remains ideal for all three applications and all $T_{confidence}$ levels.
This is because the set of configuration keywords for a given application is limited; and hence, the configuration files in the training and testing sets use the same set keywords.
Precision typically increases with the increasing $T_{confidence}$.
With a low $T_{confidence}$, the input text files with keywords that don't exist in an application vocabulary are labeled as configurations, increasing false positives.
As we show in the next section, nearly all such false positives are eliminated by checking the syntax of the selected files.
$T_{confidence}$ has a higher impact on Nginx's precision compared to httpd and MySQL as Nginx uses configuration keywords that are commonly found in the configuration files of other applications and system software (such as \texttt{user} and \texttt{include}).
To account for configuration keywords that may not be observed during training, we set $T_{confidence}$ to 0.9.
However, in our dataset, we observe no difference in the file identification results after syntax check for $T_{confidence}$ between 0.8 and 1.0.

\subsubsection{Comparison with Baselines}

We implement two baselines using the Augeas configuration editing library~\cite{Augeas} to compare \textit{ConfEx}'s configuration file identification accuracy with.
Our first baseline, \textit{default}, uses Augeas' discovery approach of checking the existence of files in specific file paths.
These paths account for the default application installation paths in various Linux distributions.
Table~\ref{tab:augeas_discovery} shows the paths checked by Augeas to identify httpd configuration files as an example.
Our second baseline, \textit{syntax}, attempts to parse all text files using Augeas, and marks the files that conform with the target application's configuration file syntax as configuration files.

\begin{table}[t]
\centering
\caption{File paths checked by Augeas to identify httpd configuration files.\newline``*'' is a wildcard that represents any file name.}
\label{tab:augeas_discovery}
\begin{tabular}{l}
\toprule
\texttt{/etc/httpd/conf/httpd.conf} \\
\texttt{/etc/httpd/httpd.conf} \\
\texttt{/etc/httpd/conf.d/*.conf} \\
\texttt{/etc/apache2/sites-available/*} \\
\texttt{/etc/apache2/mods-available/*} \\
\texttt{/etc/apache2/conf-available/*.conf} \\
\texttt{/etc/apache2/conf.d/*} \\
\texttt{/etc/apache2/ports.conf} \\
\texttt{/etc/apache2/httpd.conf} \\
\texttt{/etc/apache2/apache2.conf} \\
\bottomrule
\end{tabular}
\vspace{-0.1in}
\end{table}

\begin{figure}[t]
\subfloat[Precision\label{fig:discovery:comparison:precision}]
{\includegraphics[width=0.49\columnwidth]{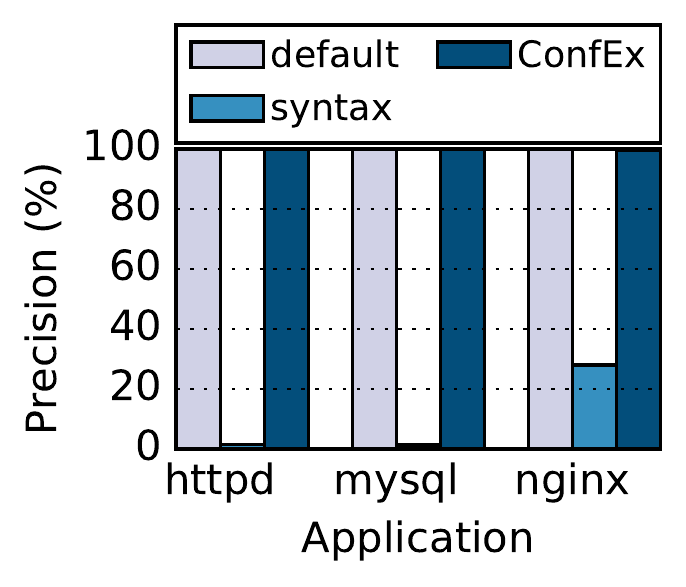}}
\subfloat[Recall\label{fig:discovery:comparison:recall}]
{\includegraphics[width=0.49\columnwidth]{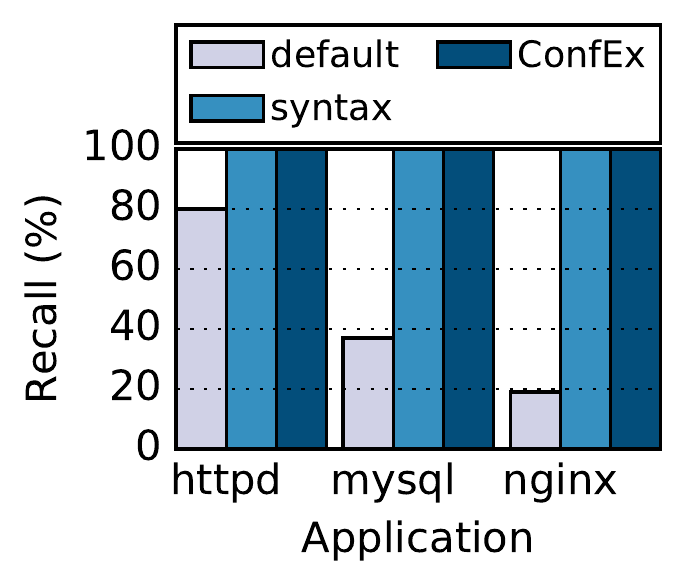}}
\caption{Configuration file identification results. While the default approach of checking standard file paths misses up to 81\% of configuration files, only checking file syntax leads to low precision by labeling non-configuration files as configurations. \textit{ConfEx}'s file labeling achieves over 99\% precision and 100\% recall.}
\label{fig:discovery:comparison}
\vspace{-0.20in}
\end{figure}

Figure~\ref{fig:discovery:comparison} compares the precision and recall of the baselines and \textit{ConfEx} on identifying application configuration files.
As all files found in default configuration paths are indeed application configurations, \textit{default} achieves ideal precision.
However, \textit{default} misses 20-81\% of configuration files, leading to low recall.
\textit{syntax} achieves ideal recall as all application configuration files conform with the expected file syntax.
However, a significant portion of non-configuration files are also syntactically valid, leading to low precision.
For httpd and MySQL, where the configuration syntax accepts key-value pairs with a space delimiter, the precision of \textit{syntax} is below 2\%.
For Nginx, where the configuration syntax involves the use of semicolons and curly brackets, the precision still remains below 29\%.

By combining the vocabulary-based check and syntactic validation, \textit{ConfEx} achieves over 99\% precision and ideal recall for all three applications.
The majority of the remaining mislabeled files are configuration files of different applications and have a single configuration command with a keyword \texttt{Include} or \texttt{include}.
These files also have the correct syntax and would be accepted as valid configurations by our target applications.

\subsection{Computational Overhead}

Crawling the contents of an image with \textit{ConfEx} takes eight seconds on the average using an Intel Xeon E5-2650 processor.
During crawling, the contents of all text files that are smaller than 200KB are compressed and recorded, excluding the files with extensions that are commonly used for non-configuration files (such as \texttt{.h} or \texttt{.css}).
Note that the contents of an image is crawled before deployment, avoiding any performance overhead on applications.
Identifying configuration files in an image and extracting the information in these files take 3.5 seconds on the average, where the mean number of text files per image is above 2800.

The only performance impact on a production system is incurred during active file discovery.
\textit{ConfEx} uses the \texttt{auditd} tool to track \texttt{open()} system calls in containers during application initialization, which is less than ten seconds in our experiments.
We observe that typically, at most a few hundred \texttt{open()} calls are issued during this initialization period, resulting in negligible performance overhead.

%% file: case_studies.tex
\vspace{-0.05in}
\section{Applied Examples}
\label{sec:case}

\textit{ConfEx} enables the use of existing key-value-based configuration analysis tools in the cloud.
To demonstrate this, we present applied examples of \textit{ConfEx} for detecting misconfigurations using two techniques proposed in prior work: \textit{PeerPressure}~\cite{Wang:OSDI04:peerPressure} and \textit{Encore}~\cite{Zhang:ASPLOS14:encore}.

\subsection{PeerPressure}
\label{sec:case:peerpressure}

\textit{PeerPressure}~\cite{Wang:OSDI04:peerPressure} is a tool that finds the culprit configuration entry in a Windows image with a single configuration error.
\textit{PeerPressure} is designed for Windows registry, where configurations are represented as key-value pairs; and hence, it is not directly applicable on text-based software configurations found in cloud instances.
In this example, we show how \textit{ConfEx} and \textit{PeerPressure} can be used together to detect misconfigurations in Docker Hub images, and study the impact of \textit{ConfEx}'s file labeling (Sec.~\ref{sec:framework:file_identification}) and disambiguation (Sec.~\ref{sec:framework:disambiguation}) steps on the effectiveness of \textit{PeerPressure}.

\textit{PeerPressure} has an offline training and an online testing phase.
During training, \textit{PeerPressure} records the histogram of values assigned to each key in a trusted configuration corpus.
Given a new image during testing, \textit{PeerPressure} compares the values assigned to each key in the image with the value histograms seen during training.
For each key-value pair, it then calculates the probability of being a misconfiguration based on empirical Bayesian estimation.
Here, an outlier value has a high probability of being an error.
Finally, the key-value pairs are ranked based on the calculated probabilities, so that the pairs that are ranked the highest are the most likely misconfigurations.

We use \textit{PeerPressure} to detect the application misconfigurations listed in Table~\ref{tab:real_misconfigs}.
We inject each misconfiguration %
to a randomly selected image that contains the target parameter to be misconfigured, and repeat the randomized injection 1000 times.
For each injection, we train \textit{PeerPressure} using the configuration key-value pairs extracted from all images except for the misconfigured image.
We then run \textit{PeerPressure} and record its output ranking for the injected error.

\begin{table}[t]
\centering
\setlength\tabcolsep{2pt}
\caption{Injected application misconfigurations}
\vspace{-0.1in}
\label{tab:real_misconfigs}
\begin{tabular}{ccl}
\toprule
application & name & \multicolumn{1}{c}{description} \Bstrut\\
\hline
httpd & url     & Error 401 points to a remote URL~\cite{misconfig_httpd_1}\Tstrut\\
httpd & dns     & Unnecessary reverse DNS lookups~\cite{misconfig_httpd_2} \\
httpd & path    & Wrong module path \\
httpd & mem     & MaxMemFree should be in KB\\
httpd & req     & Too low request limit per connection \\
MySQL & enum    & Enumerators should be case-sensitive~\cite{Xu:SOSP13:doNotBlame} \\
MySQL & buf     & Unusually large sort buffer~\cite{misconfig_mysql_2} \\
MySQL & limit   & Too low connection error limit~\cite{misconfig_mysql_3}\\
MySQL & max     & Invalid value for max \# of connections\\
Nginx & files   & Too few open files are allowed per worker\\
Nginx & debug   & Logging debug outputs to a file~\cite{misconfig_nginx_2}\\
Nginx & access  & Giving access to root directory~\cite{misconfig_nginx_4_5}\\
Nginx & host    & Using hostname in a listen directive~\cite{misconfig_nginx_4_5}\\
\bottomrule
\end{tabular}
\end{table}

We use two baselines for comparison, where we train and test \textit{PeerPressure} using the configuration key-value pairs provided by the baselines.
The first baseline, \textit{Augeas}, parses the configuration files located in default paths using Augeas library, and directly uses Augeas library's output.
The second baseline, \textit{ConfEx\_no\_disambiguation}, again uses Augeas library's output but parses all application configuration files in a given image.
We use this baseline to focus on the impact of disambiguation.

\subsubsection{PeerPressure Error Detection Results}

\begin{figure}[t]
\centering
\includegraphics[width=\columnwidth]{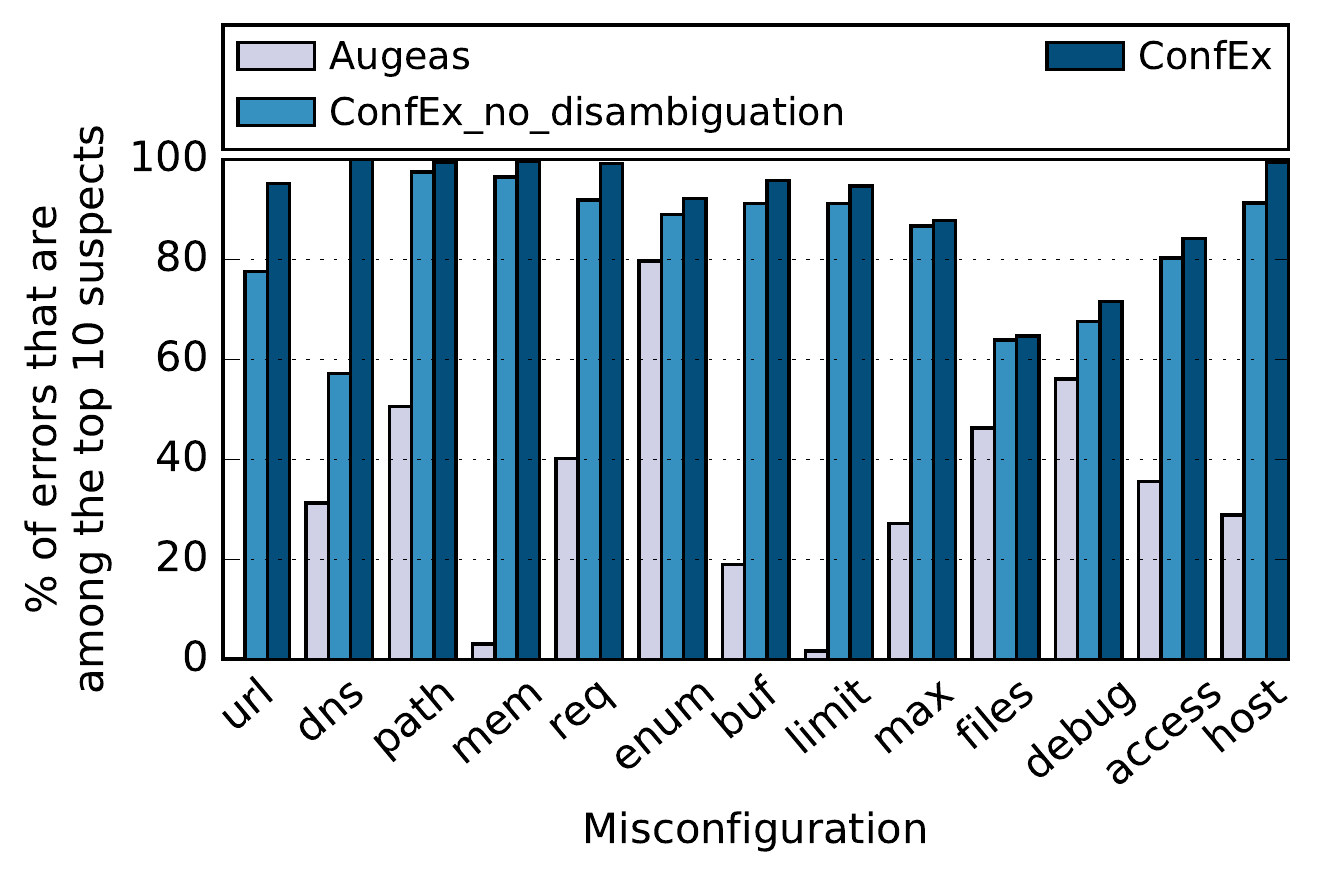}
\vspace{-0.3in}
\caption{The percentage of injected errors that are ranked within the top ten suspects by \textit{PeerPressure} among 1000 randomized injections for each misconfiguration. Using \textit{ConfEx}'s disambiguated key-value pairs consistently improves \textit{PeerPressure}'s effectiveness on error detection.%
}
\label{fig:peerpressure}
\vspace{-0.1in}
\end{figure}

Figure~\ref{fig:peerpressure} shows the percentage of injected errors that are ranked among the top 10 suspects by \textit{PeerPressure}.
Using \textit{ConfEx}'s disambiguated output consistently leads to similar or higher rankings compared to using the key-value pairs generated by baselines, making it easier to pinpoint the error.
9 out of 13 errors are ranked within the top 10 suspects for more than 90\% of the injections.

As the \textit{Augeas} baseline only parses the files located in standard paths, \textit{PeerPressure} cannot find an injected error with this baseline if the error is injected in a file in a non-standard location.
The parameters that are modified in the \textit{url}, \textit{mem}, and \textit{limit} misconfigurations rarely appear in files located in standard paths in our corpus; hence, \textit{PeerPressure} misses these errors with the \textit{Augeas} baseline.

When using the key-value pairs generated by the \textit{ConfEx\_no\_disambiguation} baseline, \textit{PeerPressure} suffers from having an incorrect view on the distribution of configurations due to the ambiguity in Augeas library's output as discussed in Sec.~\ref{sec:framework:extraction:augeas}.
This problem is exacerbated when the misconfigured image has files that have substantially different parameter ordering compared to the files seen in the corpus.
In these images, the parameters are represented by keys that use uncommon key indexing, making common configuration entries become outliers in the corpus and have high \textit{PeerPressure} rankings.

\subsection{Encore}
\label{sec:case:encore}

\textit{Encore} is a recently proposed tool that infers configuration constraints based on configurations collected from working systems~\cite{Zhang:ASPLOS14:encore}.
In its original implementation, \textit{Encore} uses the Augeas library to collect configurations and convert the collected information into key-value pairs.
As we show in this paper, using Augeas only is not sufficient to discover configuration files in cloud instances, and the key-value pairs generated by Augeas are ambiguous, decreasing the robustness of corpus-based analysis.
In this example, we demonstrate how \textit{ConfEx} significantly increases the effectiveness of \textit{Encore} on analyzing configurations in the cloud.

\textit{Encore} has two steps: Parameter type inference and rule inference.
Parameter type inference focuses on associating keys with configuration types such as integer, file path, or IP address.
To reduce the amount of generated false types and rules, \textit{Encore} focuses only on keys whose value show a certain level of \textit{entropy} %
across the corpus, where entropy is a measure of diversity.
Given a key-value pair, \textit{Encore} first checks the syntax of the value to understand the type of the configuration key.
For example, a value is identified as an IP address if it matches the regular expression \texttt{\^{}\textbackslash d\{1,3\}(\textbackslash .\textbackslash d\{1,3\})\{3\}\$}.
Next, \textit{Encore} performs a type-specific semantic check to verify the type identification.
For example, if a value is syntactically identified as a file path, the semantic step checks whether the given file path exists in the file system.
If there is such a file exists, the type of the corresponding key is inferred as a file path.

In the rule inference step, \textit{Encore} utilizes association rule learning %
to generate configuration constraints using type-specific rule templates.
An example rule template is: ``An entry should be equal to another entry of the same type''.
Such rule templates decrease the search space across the key-value pairs and reduces false rules.
For each rule template, \textit{Encore} generates all possible rules across the training corpus, then filters the generated rules based on \textit{support} and \textit{confidence}.
Support is the fraction of images that have all the keys used in the proposed rule, and confidence is the fraction of the images where the rule is valid.
\textit{Encore} accepts a rule only if its support and confidence are at least 10\% and 90\%, respectively.

We use the configuration data extracted by \textit{ConfEx} or one of the \textit{Augeas} or \textit{ConfEx\_no\_disambiguation} baselines (see Sec.~\ref{sec:case:peerpressure}) with \textit{Encore} to infer configuration parameter types and configuration value constraints.
Then, without injecting any misconfigurations, we identify the key-value pairs that violate the inferred parameter types and value constraints.

\subsubsection{Encore Error Detection Results}

Using the key-value pairs generated by our \textit{ConfEx} framework, \textit{Encore} detects 184 configuration errors.
181 of these errors are placeholder values that need to be replaced by external scripts.
The values used in these files include \texttt{\_\_PROXY\_PASS\_\_} and \texttt{\{\{(8*flavor[\textquotesingle ram\textquotesingle]/512)\textbar int\}\}M}, which would lead to errors if these files are directly used.
\textit{Encore} also detects two file path errors where Windows-style paths are used in the configuration files of Docker Hub images.
Another detected error is in the \texttt{/etc/passwd} file of an image, where the absolute path of a user's shell is set to \texttt{ata:ata:at}.

When using the \textit{ConfEx\_no\_discovery} baseline, \textit{Encore} finds only 99 of these errors due to the ambiguity in the key-value pairs.
Using the \textit{Augeas} baseline, which ignores files in non-standard paths both during training and testing, reduces the number of detected errors to 10.

%% file: related.tex
\section{Related Work}
\label{sec:related}

Finding and preventing errors is a major focus of the research on software configurations.
Execution trace analysis and binary instrumentation have been shown to provide insight on the root causes of configuration errors~\cite{Attariyan:OSDI12:xray,Zhang:ICSE14:whichConfig}.
Due to their intrusiveness, however, instrumentation and trace analysis are often impractical on production workloads.
Source code analysis~\cite{Xu:SOSP13:doNotBlame,Zhou:EASE17:easier,%
Nadi:ICSE14:mining} and natural language processing on application documentations~\cite{Potharaju:VLDB15:confSeer,Jin:ASE14:prefFinder} have been used to infer configuration constraints before deployment.
Configuration entries that do not comply with these constraints are then marked as errors. 

Runtime-based configuration validation techniques~\cite{jahanbanifar:ISORC15:partial,akue:CNSM2013:checker} have been used to check that the runtime behavior of an application matches an expected behavioral profile based on a valid configuration. Unexpected behavior is flagged as a possible misconfiguration. These techniques are largely application-specific and require either precise configuration of a monitoring daemon for the kinds of behavior being observed or modification of the application itself.

Application-agnostic statistical techniques~\cite{Wang:OSDI04:peerPressure,Zhang:ASPLOS14:encore,Santolucito:plang17:synthesizing} use previously-observed configurations to learn about the common patterns, and identify deviations as potential errors.
These techniques require key-value pairs that represent configurations for analysis, and do not address discovery or extraction for text-based configuration files.

In prior studies, the discovery and extraction of configurations in the cloud have been performed using several methods: parsing known configuration files with custom scripts (e.g.,~\cite{chieu:ICEE12:cloud,Potharaju:VLDB15:confSeer}), crawling erroneous files from mailing lists and technical forums (e.g.,~\cite{Xu:OSDI16:earlyDetection}), parsing files located in default paths using configuration parsing libraries (e.g.,~\cite{Zhang:ASPLOS14:encore}), and using standardized configuration stores such as the Windows registry (e.g.,~\cite{Wang:OSDI04:peerPressure}).
In image repositories and multi-tenant cloud environments, however, configuration file locations are unknown, and configuration parsers produce key-value pairs that lack the consistency and robustness required for meaningful statistical analysis.

Existing tools for handling configurations focus on centralized management rather than extracting key-value pairs in a cloud environment.
Chef~\cite{chef} and Ansible~\cite{Ansible} have configuration editing capabilities that are restricted to search-and-replace based on regular expressions, but they cannot extract configurations from text files.
CFEngine~\cite{CFEngine} can parse standard file formats such as XML and JSON, but not application-specific configuration formats such as in httpd and Nginx.
Puppet~\cite{puppet} and bcfg2~\cite{bcfg2} can edit application-specific files by leveraging Augeas library~\cite{Augeas}.
As we show in this work, using the Augeas library alone is not sufficient for parameter extraction for robust configuration analysis.

Recently, Huang et al. proposed SAIC~\cite{Huang:arxiv17:SAIC}, a tool to help users discover text-based configuration files in cloud instances with unlabeled content.
To identify configuration files, SAIC analyzes the change patterns of files over the lifetime of a cloud instance.
Hence, SAIC is only applicable to cloud instances that have multiple versions where the configuration file locations remain the same and configurations are modified.
Xu et al.~\cite{Xu:arxiv:mining} have discussed the opportunities and challenges associated with mining container image repositories for software configurations, but have not devised a solution.

We have introduced a preliminary version of \textit{ConfEx} in our recent work~\cite{Tuncer:DSN18:confex}.
As opposed to this preliminary version, our framework in this paper has an active file discovery methodology, applies syntax check while identifying configuration files in addition to the vocabulary-based comparison, and records the order of parameters that appear in a configuration file as well as environmental configuration information.
These improvements enable the detection of a wider set of misconfigurations.
In addition to the above improvements, in this extended paper, we present a more detailed evaluation of \textit{ConfEx} by using both \textit{PeerPressure} and \textit{Encore} as applied examples.

To the best of our knowledge, our configuration analytics framework, \textit{ConfEx}, is the first to systematically discover and extract text-based software configurations in cloud instances with unlabeled content. In this way, \textit{ConfEx} enables comprehensive analysis of software configurations in the cloud to help users validate their configurations and achieve robust operation.

%% file: conclusion.tex
\section{Conclusion}
\label{sec:conclusion}

Due to the increasing prevalence and severity of misconfigurations in cloud services, there is need for a cloud-based framework that can support analyzation and validation of software configurations.
To enable automated configuration analysis, we have proposed \textit{ConfEx}, a framework to discover and analyze text-based software configurations in multi-tenant cloud platforms.
To identify configuration files in cloud instances with unlabeled content, \textit{ConfEx} keeps track of configuration keywords such as parameter names and commands in text files and checks file syntax, achieving over 99\% precision and 100\% recall.
To parse these files, \textit{ConfEx} leverages a community-driven configuration parser, Augeas.
It then disambiguates the parser output to obtain configuration key-value pairs that are consistent across different files and cloud instances.
Our framework enables the use of existing configuration analysis tools, which are designed for key-value pairs, in the cloud, which we hope will lead to fewer configuration-related outages and a more secure and reliable cloud ecosystem.

\section*{Acknowledgement}
\label{sec:ack}
This project has been partially funded by the IBM T.J. Watson Research Center.